# Construction of topological quantum magnets from atomic spins on surfaces

Hao Wang[1,2,†], Peng Fan[1,2,†], Jing Chen[1,2,†], Lili Jiang[2], Hong-Jun Gao[1,2], Jose L. Lado[3]*, Kai Yang[1,2]*

[1]Beijing National Laboratory for Condensed Matter Physics and Institute of Physics, Chinese Academy of Sciences, Beijing 100190, China

[2]School of Physical Sciences, University of Chinese Academy of Sciences, Beijing 100049, China

[3]Department of Applied Physics, Aalto University, 02150 Espoo, Finland

*Corresponding authors: K.Y. (kaiyang@iphy.ac.cn); J.L.L. (jose.lado@aalto.fi)

†These authors contributed equally to this work.

**Artificial quantum systems have emerged as indispensable platforms to realize exotic topological matter in a well-controlled manner. Here, we demonstrate topological quantum Heisenberg spin lattices, engineered with spin chains and two-dimensional spin arrays using spin 1/2 atoms on insulating films in a scanning tunnelling microscope (STM). We engineered with atomic precision both topological and trivial phases of the quantum spin model, realizing first- and second-order topological quantum magnets. Their many-body excitations were probed by single-atom electron spin resonance with ultrahigh energy resolution. The atomically-localized magnetic field of the STM tip allows us to directly visualize various topological bound modes including topological edge states, topological defects, and higher-order corner modes. Our results provide an important bottom-up approach to simulating exotic quantum many-body phases of interacting spins.**

The exploration of topological properties in condensed matter systems has sparked a paradigm shift in our comprehension of quantum phenomena, promoted for their potential for protected quantum information processing [1,2]. Recently, the development of quantum simulation has allowed the realization of topological matter in a well-controlled manner in several platforms such as semiconductor quantum dots [3] and cold atoms [4,5]. For example, the many-body Su-Schrieffer-Heeger (SSH) model has been studied in precision-placed donors in silicon [3] as well as using Rydberg atoms [4].

Scanning tunnelling microscope (STM) can be used to fabricate precisely-engineered topological matter at the atomic scale in a solid-state environment [6], such as single-particle SSH dimer chains [7] and electronic higher-order topological insulators [8]. In addition to the charge degree of freedom, spin lattices constructed on surfaces has emerged as an attractive atomic-scale platform for quantum simulations of exotic phases of matter [9-14], including Yu-Shiba-Rusinov lattices [13] and resonating valence bond states [14]. However, it remains a formidable task to fabricate and sense topological quantum magnets with atomic precision. Recently, effective spin-1 Haldane chains composed of ferromagnetically-coupled spin-1/2 radicals were built using organic molecules on Au(111), where the spin-1/2 end states are Kondo-screened by the conduction electrons from the substrate [15,16]. While previous studies have focused on one-dimensional (1D) topological spin chains on metals, topological spin chains or two-dimensional (2D) spin arrays have never been fabricated on a decoupling layer that protects the spins from screening, despite its vital importance in demonstrating their intrinsic topological behavior—mostly owing to the challenge of precisely engineering magnetic interactions between atoms on insulators to reproduce the required spin Hamiltonian.

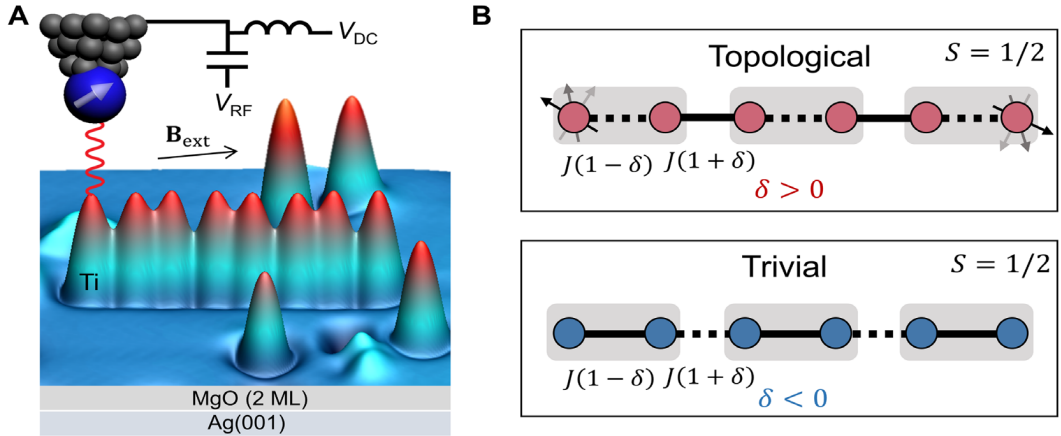

**Fig. 1. Realization of dimerized antiferromagnetic spin-1/2 lattices on a surface.** (**A**) Experimental setup showing an STM with ESR capability, and an STM image of Ti atoms on MgO/Ag(001). A spin-polarized tip is used to drive and sense ESR. (**B**) Dimerized spin-1/2 chains with alternating antiferromagnetic coupling strengths $J(1-\delta)$ and $J(1+\delta)$, showing topological phase ($\delta > 0$, upper panel) and trivial valence-bond solid phase ($\delta < 0$, lower panel). Grey rectangles indicate the chosen unit cells.

Here, we constructed topological quantum spin lattices including 1D spin chains and a 2D spin array using spin-1/2 Ti atoms on MgO in a low-temperature STM (Fig. 1). By finely tuning the antiferromagnetic exchange interaction, both topological and trivial phases are realized (Fig. 1B). Combining electron spin resonance (ESR) with STM, we measured their many-body spin transitions atom-by-atom with an ultrahigh energy resolution below 100 neV. Atomic-scale magnetic resonance imaging of these engineered spin lattices allows us to directly visualize the topological edge states and topological defects in the spin chains, as well as the higher-order corner modes in the 2D spin array. Our theoretical modeling shows that these localized topological modes originate from a non-trivial Berry phase, providing crucial insight into the topological properties of quantum Heisenberg spin lattices.

The spin lattices were made by positioning Ti atomic spins on the insulating two-monolayer MgO film grown on Ag(001) [14,17,18]. For both adsorption sites of Ti on MgO, the Ti atom has spin $S = 1/2$ with no single ion magnetocrystalline anisotropy. Two Ti spins couple via antiferromagnetic exchange interaction at close distance (< 1 nm), as revealed by single-atom ESR [17,18]. The exchange coupling constant $J$ shows exponential dependence on the separation $r$ between Ti atoms, as described by an exponential function, $J = J_0 \exp(-(r-r_0)/d)$ where $r_0$ is an arbitrary reference distance. For two bridge-site Ti atoms, $J_0$ = 27.7±0.1 GHz, $d$ = 0.94±0.01 Å and $r_0$ = 7.2 Å [14,18]. We fabricated spin lattices using bridge-site Ti atoms in the following.

**1D topological quantum magnet**

We first focus on the 1D topological spin model, which is realized via an engineered dimerization in a spin-1/2 chain. Their quantum states under external applied magnetic field $B_{ext}$ (almost in-plane) are described by the following spin Hamiltonian [19]:

$$H = J(1-\delta)\sum_n^{L/2} \mathbf{S}_{2n-1}\cdot\mathbf{S}_{2n} + J(1+\delta)\sum_n^{L/2-1}\mathbf{S}_{2n}\cdot\mathbf{S}_{2n+1} + \sum_n^L g\mu_B \mathbf{B}_{ext}\cdot\mathbf{S}_n + g\mu_B\mathbf{B}_{tip}\cdot\mathbf{S}_m \quad (1)$$

where $\mathbf{S}_n = (S_n^x, S_n^y, S_n^z)$ is the spin operator for site $n$ and the total number of sites is $L$. The exchange coupling $J$ is 15.5 GHz and dimerization constant $\delta$ is about 0.61 for the spin chains constructed, corresponding to interatomic separations of 0.86 nm and 0.73 nm (3×0 and 2.5×0.5 lattice constants of MgO). The g-factor $g \approx 1.8$ is obtained by measuring ESR of isolated Ti atoms, and $\mu_B$ is the Bohr magneton. Since the exchange coupling is exponential in distance, second-nearest-neighbor coupling can be neglected. The direction of the external field $B_{ext}$ is defined as $z$ in the following. The atomic-scale tip magnetic field $B_{tip}$ is used both to drive ESR transitions and to tune the many-body spin states by exerting an exchange bias only on the spin $S_m$ under the tip [20].

In single-particle SSH model, there exists a symmetry-protected topological phase which gives rise to two zero-energy edge states where the electron is localized at the edge sites of the lattice. Interestingly, a quantum-many-body generalization based on dimerized quantum spin-1/2 model (eq. 1) also features topological edge modes, stemming from the existence of a non-trivial Berry phase of a fractional pseudo-fermionic representation (Supplementary Sec. 2). For the dimerized quantum spin chains on MgO, the edge topological spin excitations can be accessed by means of a weak external magnetic field as shown below. It is important to note that, for sufficiently large chains, the dimerized Heisenberg model has a ground state that is four-fold degenerate when $\delta > 0$, consisting of a singlet and a triplet state, resulting from the dangling edge excitations.

**Topological edge modes of 6- and 8-spin chains**

To realize these topological spin edge states, we first built a 6-spin chain in the topological configuration ($\delta > 0$) with alternating interatomic distances to obtain nearest-neighbor coupling of 6 and 25 GHz (Fig. 2A). These coupling strengths allow transitions between spin multiplets to be visible in our ESR range of ~13–23 GHz. Each spin multiplet with a total spin $S_T$ fans out into its $2S_T + 1$ components when $B_{ext}$ is increased (Fig. S8), as expected for Heisenberg spin Hamiltonian. When $B_{ext}$ = 0, the two lowest multiplets are one spin singlet and one spin triplet, resulting from the effective exchange coupling between the two edge states. With increasing $B_{ext}$, the triplet state aligned with the field becomes the ground state.

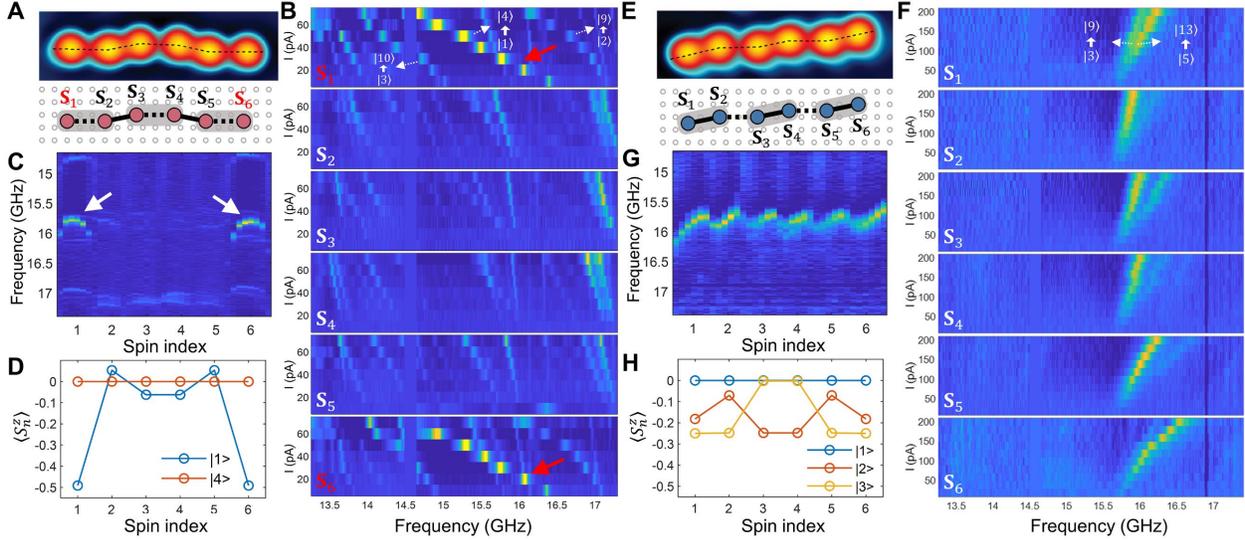

**Fig. 2. Topological versus trivial configurations of 6-spin chains.** (**A**) STM image (1.8 nm × 6 nm) and binding sites for the topological configuration (grey circles are oxygen sites of MgO). (**B**) ESR spectra measured on each of the 6 spins as a function setpoint current $I$, which is approximately proportional to $B_{tip}$ ($V_{DC}$ = 60 mV, $I$ = 10–70 pA, $V_{RF}$ = 18–35 mV, $B_{ext}$ = 0.69 T). Red arrows indicate the edge modes. (**C**) ESR spectra measured along dashed line in (A) ($V_{DC}$ = 60 mV, $I$ = 20 pA, $V_{RF}$ = 32 mV). White arrows indicate the edge modes. (**D**) Calculated average magnetization $\langle S_n^z \rangle$ of the topological configuration for its ground state $|1\rangle$ and excited state $|4\rangle$. (**E**) STM image (1.7 nm × 5.5 nm) and binding sites for a trivial configuration. (**F**) ESR spectra measured on each of the 6 spins ($V_{DC}$ = 50 mV, $I$ = 20–200 pA, $V_{RF}$ = 16–36 mV, $B_{ext}$ = 0.61 T). (**G**) ESR spectra measured along dashed line in (E) ($V_{DC}$ = 50 mV, $I$ = 100 pA, $V_{RF}$ = 28 mV). (**H**) Calculated $\langle S_n^z \rangle$ of the trivial configuration for its ground state $|1\rangle$ and excited states $|2\rangle$, $|3\rangle$.

To probe the eigenstates of this 6-spin chain, we performed ESR on different sites and studied the evolution of the ESR spectra as a function of $B_{tip}$ (Fig. 2B). Since different ESR transitions respond differently with increasing $B_{tip}$, this allows us to easily identify the corresponding initial and final states, in accordance with the quantitative simulations. This also allowed us to extract information about the spin wavefunction including spin polarization on different sites. The ESR peaks as indicated by red arrows in Fig. 2B reveal the transition from the ground state $|1\rangle$ to the excited state $|4\rangle$, which is the transition between the low-energy spin triplet ($|1\rangle = |S_T = 1, m_z = -1\rangle$) and singlet ($|4\rangle = |S_T = 0, m_z = 0\rangle$). This transition is visible only when the tip is positioned at either end of the chain (i.e., spins $S_1$ or $S_6$), and its frequency shifts linearly with increasing $B_{tip}$. Since all spins have almost zero polarization in state $|4\rangle = |m_z = 0\rangle$, this shows that the spin distributions of the ground state $|1\rangle$ is mainly localized on the two end spins. These localized ESR transitions are attributed to the edge topological spin excitations. The localization of the ground state is similar to the SSH model, and this spin edge state can be intuitively understood as a many-body version of a symmetry-protected topological state.

To visualize the spatial distribution of the topological edge modes, we measured the ESR spectra along the spin chain as indicated by the dashed line in Fig. 2A. The spatial mapping of the ESR spectra shown in Fig. 2C clearly reveals the localization of the spin distribution to the edges of the chain in the ground state $|1\rangle$. The triplet state with $m_z = -1$ is the ground state under external magnetic field, giving rise to a finite magnetization on the edges. We calculated the local average magnetization $\langle S_n^z \rangle$ on each site using the ground state $|1\rangle$ and the excited state $|4\rangle$. The localization of the ESR transition $|1\rangle \to |4\rangle$ observed in the experiment agrees well with the spatial distribution of $\langle S_n^z \rangle$ as shown in Fig. 2D.

For comparison, we also constructed a dimerized 6-spin chain with the trivial configuration ($\delta < 0$) by finely tuning the antiferromagnetic exchange interaction (Fig. 2E). Its ground state is a spin singlet, corresponding to a valence-bond solid configuration, which is further from excited states in the energy-level diagram than the $\delta > 0$ configuration (Fig. S8). In contrast to the strongly localized ESR transitions observed for the $\delta > 0$ configuration, ESR spectra and ESR mapping measured on the trivial one display almost uniform spatial distribution along the atomic chain (Fig. 2, F and G), indicating the absence of topological edge modes.

In addition to these strong ESR peaks localized at the end spins, several weaker ESR peaks are also visible on all spins in the chains (Fig.2, B and F). These peaks correspond to transitions between other many-body spin states (as indicated in the top panels of Fig. 2, B and F), providing rich information about the excited states of the dimerized spin chains.

We further compare the topological and trivial configurations of the dimerized spin model by increasing the number of atomic spins to eight (Fig. 3 and Fig. S11). With increasing number of atoms, the low-energy edge triplet and singlet are expected to get closer in energy, and the local average magnetization $\langle S_n^z \rangle$ should decay exponentially with $n$ for an infinite spin chain. For the topological configuration of the 8-spin chain, atomic-scale ESR spectra exhibit localized ESR transitions at the two end spins, corresponding to the edge topological spin excitations occurring within the low-energy edge triplet (Fig. 3B and Fig. S9). In addition, ESR spectroscopic mapping measured along the chain also displays strong localization on either end of the chain, as indicated by white arrows in Fig. 3C. The calculated $\langle S_n^z \rangle$ shows dominant weight on the two end spins due to the triplet ground state, which causes the observed localized ESR transitions (Fig. 3D). Similar to the 6-spin chain, many other weaker ESR peaks are also visible in the ESR spectra measured on all spins of the 8-spin chain.

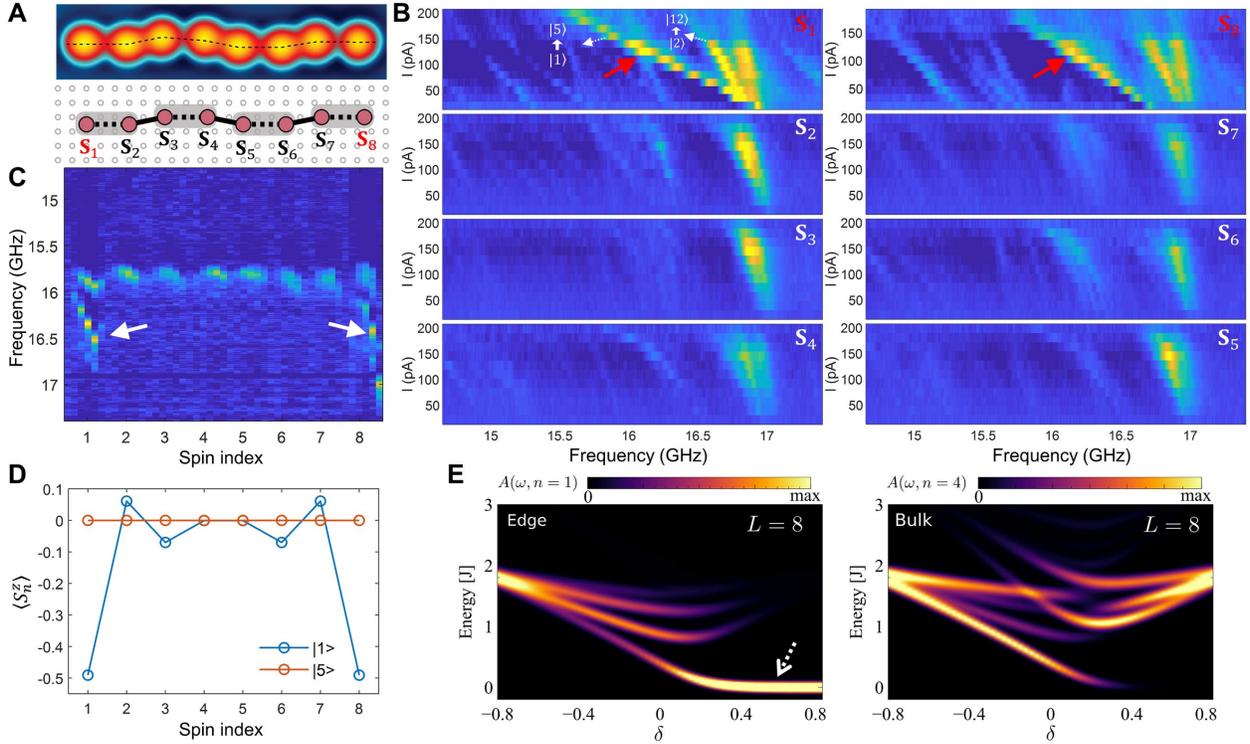

**Fig. 3. Topological edge states of an 8-spin chain.** (**A**) STM image (1.5 nm × 7 nm) and binding sites of a dimerized 8-spin chain on MgO with topological configuration. (**B**) ESR spectra measured on each of the 8 spins as a function of setpoint current $I$, which is approximately proportional to $B_{\text{tip}}$ ($V_{\text{DC}}$ = 50 mV, $I$ = 20–200 pA, $V_{\text{RF}}$ = 9–29 mV, $B_{\text{ext}}$ = 0.67 T). Red arrows indicate the edge modes. (**C**) ESR spectra measured along the dashed line in (A) ($V_{\text{DC}}$ = 50 mV, $I$ = 120 pA, $V_{\text{RF}}$ = 22 mV, $B_{\text{ext}}$ = 0.61 T). White arrows indicate the edge modes. (**D**) Calculated average magnetization $\langle S_n^z \rangle$ for its ground state $|1\rangle$ and excited state $|5\rangle$. (**E**) Evolution of the spin spectral function $A(\omega, n)$ as a function of $\delta$. For $\delta > 0$, a zero-energy edge mode appears (left, white arrow) and bulk spectral gap opens (right).

## Topological origin of edge spin modes

Figure 3E shows the calculated many-body spin spectral function as a function of the dimerization constant $\delta$, which indicates the emergence of edge modes for $\delta > 0$. The spectral function provides a model of the observed excitations. To better elucidate the topological origin of the observed spin-1/2 localized edge modes in the dimerized spin chains, the dimerized Heisenberg model [ $H = J(1-\delta)\sum_n^{L/2} \mathbf{S}_{2n-1} \cdot \mathbf{S}_{2n} + J(1+\delta)\sum_n^{L/2-1} \mathbf{S}_{2n} \cdot \mathbf{S}_{2n+1}$ ] is adiabatically transformed to a dimerized pseudo-fermion model of the form (Supplementary Sec. 2):

$$H = t_1 \sum_n^{L/2} f_{2n-1}^\dagger f_{2n} + t_2 \sum_n^{L/2-1} f_{2n}^\dagger f_{2n+1} + h.c. \quad (2)$$

where $f_n$ is a pseudo-fermion operator and $t_1/t_2 = (1-\delta)/(1+\delta)$. The previous Hamiltonian can be solved in reciprocal space, and hosts a non-trivial Berry phase $\phi = \pi$ for $t_1/t_2 < 1 (\delta > 0)$, defined by $\phi = i \oint \langle \Psi_k | \partial_k | \Psi_k \rangle dk$, with $\Psi_k$ the Bloch eigenfunctions. Thus, within this mapping to fractional pseudo-fermions, the existence of the non-trivial Berry phase of the pseudo-fermion model gives rise to the observed topological edge modes of the original dimerized spin Hamiltonian [21].

## Topological defect in a dimerized 9-spin chain

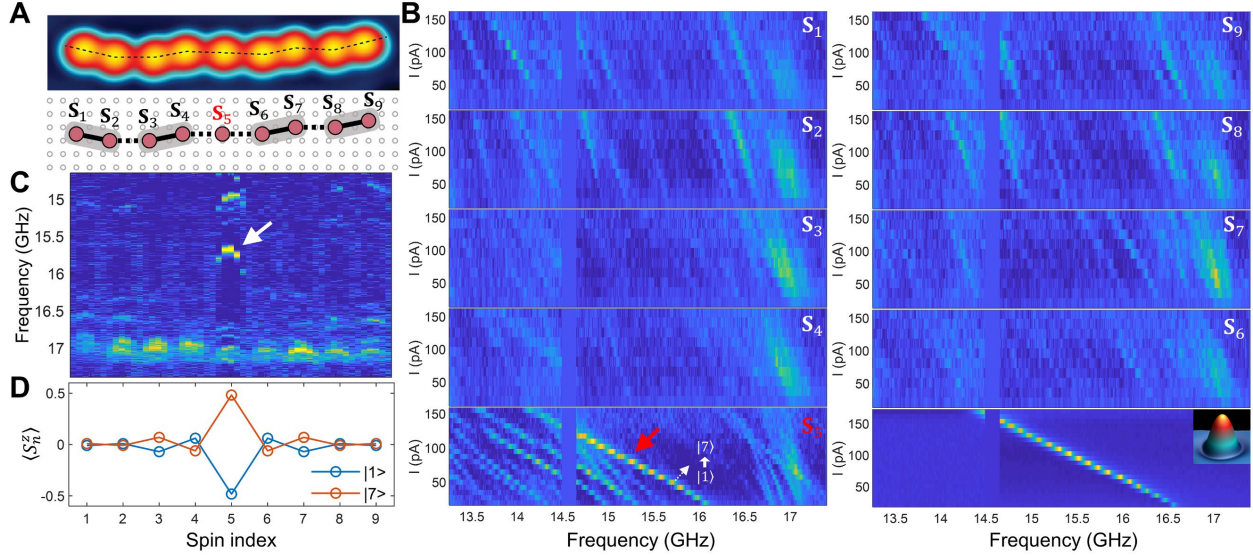

**Fig. 4. Topological mode of a 9-spin chain.** (**A**) STM image (1.75 nm × 8.3 nm) and binding sites of a dimerized 9-spin chain on MgO with a topological defect in the middle. (**B**) ESR spectra measured on each of the 9 spins as a function of setpoint current $I$, which is approximately proportional to $B_{tip}$ ($V_{DC}$ = 50 mV, $I$ = 20–155 pA, $V_{RF}$ = 6–30 mV, $B_{ext}$ = 0.68 T). The red arrow indicates the bound mode. Lower right panel: ESR spectra measured on a single Ti spin ($V_{DC}$ = 50 mV, $I$ = 20–170 pA, $V_{RF}$ = 6–30 mV). (**C**) ESR spectra measured along the dashed line in (A) ($V_{DC}$ = 50 mV, $I$ = 50 pA, $V_{RF}$ = 20 mV). The white arrow indicates the bound mode. (**D**) Calculated average magnetization $\langle S_n^z \rangle$ for its ground state $|1\rangle$ and excited state $|7\rangle$.

In addition to dimerized spin chains with topological edge states as shown above, we also fabricated a dimerized 9-spin chain with a topological defect in the middle (Fig. 4A). The middle spin can be viewed as the domain boundary between two different dimer configurations. This would create a topological bound mode in the middle of chain. To verify this, we measured the ESR spectra along the chain and observed an additional strong ESR transition only on the middle spin (as indicated by the red arrow in the lower left panel of Fig. 4B). With increasing $B_{tip}$ that is atomically localized at the middle of the chain, the frequency of this ESR peak shifts with a similar rate as the ESR peak measured on an isolated Ti spin (Fig. 4B, lower right panel). This suggests that the local spin polarization of the topological mode is similar to an isolated spin, which agrees well with our theoretical calculation of the local average magnetization $\langle S_n^z \rangle$ (Fig. 4D). The spin-1/2 topological defect can also be clearly recognized in the ESR mapping in Fig. 4C.

## Higher-order topological quantum magnet

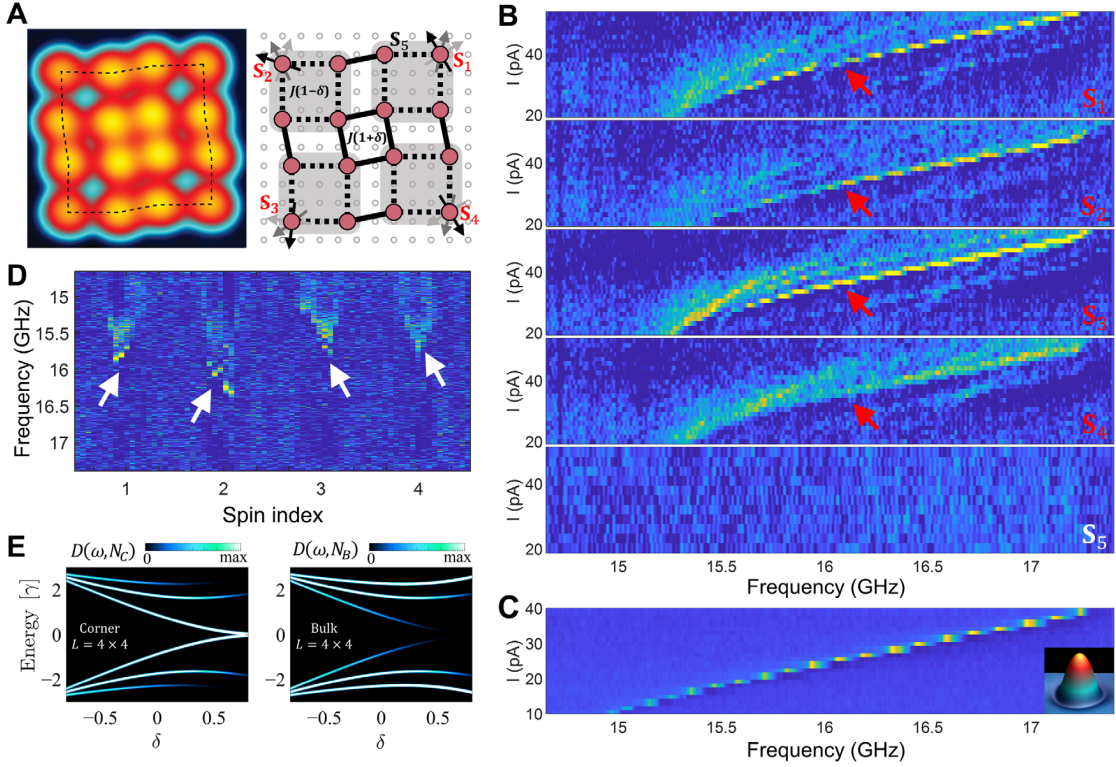

**Fig. 5. Higher-order topological mode of a 4×4 spin lattice.** (**A**) STM image (4 nm × 4 nm) and binding sites of a 4×4 spin lattice. (**B**) ESR spectra measured on the corner spins ($S_1, S_2, S_3, S_4$) and on $S_5$ as a function of setpoint current *I*, which is approximately proportional to $B_{\text{tip}}$ ($V_{\text{DC}}$ = 50 mV, *I* = 20–53 pA, $V_{\text{RF}}$ = 17–30 mV, $B_{\text{ext}}$ = 0.59 T). Red arrows indicate the corner modes. (**C**) ESR spectra measured on a single Ti spin ($V_{\text{DC}}$ = 50 mV, *I* = 10–40 pA, $V_{\text{RF}}$ = 15–30 mV). (**D**) ESR spectra measured along the dashed line indicated in (A) ($V_{\text{DC}}$ = 50 mV, *I* = 32 pA, $V_{\text{RF}}$ = 26 mV). White arrows indicate the corner modes. (**E**) Parton spectral function of the 4×4 spin lattice at the corner (left) and the bulk (right). For $\delta > 0$, a corner mode appears whereas the bulk remains gapped.

We now move to the 2D spin lattice featuring higher-order topological modes. Similar to the higher-order topological insulator with topological states emerging in two dimensions lower than the bulk [22,23], the 2D dimerized spin lattice we built exhibits zero-dimensional spin excitations. Figure 5A shows a 2D dimerized 4×4 spin lattice, which can be viewed as a many-body generalization of a second-order topological insulator. By engineering alternating weak and strong coupling, we show that a topological state emerges at the corners of the square lattice. We first focus on the ESR spectra measured on the four corner spins, which show clear topological modes that appear as the strongest ESR excitation (red arrows in Fig. 5B). In contrast, the localized spin mode is absent at the edge or inner spins of the lattice (Fig 5B lower panel; Fig. S12). Importantly, other many-body spin transitions are also visible in the ESR spectra in Fig. 5B, highlighting the many-body nature of the observed topological corner modes. In addition, we also acquired the ESR spectra along the outer spins of the square lattice as indicated by the dashed line in Fig. 5A. The resulting ESR mapping is shown in Fig. 5D, which again reveals the localization nature of the corner modes.

The topological origin of the many-body corner modes can be understood by employing an auxiliary parton pseudo-fermionic representation of the original dimerized spin Hamiltonian (Supplementary Sec. 2). The emergence of topological modes can be observed by computing the parton spectral function $D(\omega, n)$. As shown in Fig. 5E, in-gap excitations appear for $\delta > 0$ at the corner of the lattice, which corresponds to the topological regime of the model. Such corner modes coexist with a gap in bulk of the lattice, and in the thermodynamic limit lead to a 16-fold topological degeneracy of the lattice.

**Outlook**

Our work shows that topological quantum magnets can be realized with spin-1/2 centers with carefully designed topology [5,19,24]. Our results establish a proof-of-concept demonstration of the potential of engineered spin lattices for realizing quantum spin liquid phases and topological order [25]. In contrast with bulk quantum magnets, atomically engineered quantum magnets probed by single-atom ESR could serve as versatile solid-state analog quantum simulators [3,26,27] because they can be assembled, modified and probed in-situ with single-spin selectivity. This technique, in combination with pump-probe electronic pulses, could allow exploration of the quantum dynamics of quasiparticles in artificial spin structures [28,29]. Ultimately, extending the size of the spin structures and introducing other degrees of freedom, such as superconductivity, would allow the exploration of exotic phases combining quantum magnetism and Yu-Shiba-Rusinov lattices [13].

## Acknowledgments

This work is supported by the National Key R&D Program of China (2022YFA1204100), the Beijing Natural Science Foundation (Z230005), the National Natural Science Foundation of China (12174433, 52272170), and the CAS Project for Young Scientists in Basic Research (YSBR-003). J.L.L. acknowledges the computational resources provided by the Aalto Science-IT project, the financial support from the Academy of Finland Projects (No. 331342 and No. 358088) and the Jane and Aatos Erkko Foundation.

## Author contributions

K.Y. and J.L.L. designed the experiment. H.W., P.F., J.C. and K.Y. carried out the STM measurements. J.L.L. developed the theoretical model. H.W., K.Y. and J.L.L. performed the analysis and wrote the manuscript with help from all authors. All authors discussed the results and edited the manuscript.

## Competing interests

The authors declare no competing interests.

## Data availability

The data that support the plots within this paper and other findings of this study are available from the corresponding authors upon reasonable request.